\documentclass[12pt]{article}

\usepackage{amsmath}
\usepackage{amssymb}
\usepackage{graphicx}
\usepackage{bm}

\usepackage{epsfig}
\usepackage{graphics}

\newcommand{\la}{\lambda}

\newcommand{\br}{{\bf r}}
\newcommand{\prt}{\partial}
\newcommand{\bu}{{\bf u}}

\begin{document}

\title{
Spatial dispersive shock waves generated in supersonic flow of
Bose-Einstein condensate past slender body}

\author{
G.A. El$^{1}$ and A.M. Kamchatnov$^{2}$\\
$^1$Department of Mathematical Sciences, Loughborough University\\
Loughborough LE11 3TU, UK \\
$^2$Institute of Spectroscopy, Russian Academy of Sciences,\\
Troitsk 142190, Moscow Region, Russia }

\maketitle

\begin{abstract}
Supersonic flow of Bose-Einstein condensate past macroscopic
obstacles is studied theoretically. It is shown that in the case
of large obstacles the Cherenkov cone transforms into a stationary
spatial shock wave which consists of a number of spatial dark
solitons. Analytical theory is developed for the case of obstacles
having a form of a slender body. This theory explains
qualitatively the properties of such shocks observed in recent
experiments on nonlinear dynamics of condensates of dilute alkali
gases.
\end{abstract}

In usual compressible fluid dynamics, there are two typical
situations when shock waves can be generated. The first one is
connected with breaking of a nonlinear wave and the second with a
supersonic flow past a body (see, e.g., \cite{LL6, whitham,
karpman}). In a viscous fluid, the shock wave can be represented
as a narrow region within which strong dissipation processes take
place and the thermodynamic and hydrodynamic parameters of the
flow undergo sharp change. If viscosity is negligibly small
compared with dispersion effects, the shock discontinuity is
resolved into an expanding region filled with nonlinear
oscillations. A remarkable feature of such a ``dispersive shock''
is generation of solitons at one of its boundaries   so that the
whole structure can often be asymptotically represented as a
``soliton train''. The theory of dispersive shocks based on the
Whitham nonlinear modulation theory was developed for media
described by the Korteweg-de Vries (KdV) equation as for the wave
breaking case \cite{GP73}, so for the flow past a slender body
\cite{GKKE95}. In the latter case the ``soliton train'' represents
a ``fan'' of oblique spatial solitons spreading downstream from
the pointed part of the body.

After experimental discovery of the Bose-Einstein condensate (BEC) \cite{bec1,bec2,bec3},
it was found that its dynamics is described very well by the Gross-Pitaevskii (GP)
equation (see, e.g., \cite{ps2003})
\begin{equation}\label{eq1}
   i\hbar\frac{\prt \psi}{\prt t}=-\frac{\hbar^2}{2m}\Delta\psi+U(\br)\psi
   +g|\psi|^2\psi,
\end{equation}
where $\psi(\br)$ is the order parameter (``condensate wave function''), $U(\br)$ is
the potential which confines atoms of Bose gas in a trap, and $g$ is an effective
coupling constant arising due to inter-atomic collisions with the $s$-wave
scattering length $a_s$,
\begin{equation}
   g=4\pi\hbar^2a_s/m,
\end{equation}
$m$ being the atomic mass.  The GP equation (\ref{eq1})  combines
the dispersive and nonlinear effects, and the corresponding
properties of BEC dynamics have been investigated in a number of
papers (see for review, e.g., \cite{ps2003}). In particular, if
$g>0$, then existence of dispersive shocks can be expected, their
theory was developed in \cite{kgk04,damski} and they were observed
in a recent experiment \cite{exp}. Although in the experiment
\cite{exp} the shock flow was strongly disturbed by vortices
arising due to rotation of the condensate, we were informed
\cite{private} about unpublished results of experiments on shocks
in non-rotating BEC, and these results agree qualitatively with
the theory \cite{kgk04}. In experiments \cite{exp,private}, the
shocks were generated by sharp push of cylindrical BEC from its
axis by a laser beam propagating along the axis. After breaking of
a cylindrical nonlinear wave the dispersive shock occurs which
consists of a train of concentric cylindrical solitons.

Besides the experiments on generation of shocks after wave
breaking of BEC's flow, in \cite{private} the experiments were
performed on BEC's flow past an obstacle after release of BEC from
the confining potential. The problem of superflow past a body has
been studied intensively in connection with a problem of critical
velocity $v_c$ above which superfluidity disappears (see, e.g.,
\cite{feynman, NP}). It was found that superfluidity is lost due
to generation of vortex rings behind an obstacle which gives the
estimate of critical velocity
\begin{equation}
   v_c\sim({\hbar}/{dm})\ln(d/\xi),
\end{equation}
where $\xi=\hbar/\sqrt{2mn_0g}$ is the healing length (i.e.,
``vortex core size''), $d$ is the size of obstacle in transverse
direction, and $n_0$ is the density of atoms in the condensate far
from the obstacle. For large obstacles with $d\gg\xi$ this
estimate gives the critical velocity much less than the sound
velocity
\begin{equation}
c_s=\hbar/\sqrt{2}m\xi.
\end{equation}
 This transition to dissipative behaviour
in quantum fluids described by the GP equation (\ref{eq1}) was
confirmed by numerical study \cite{FPR,WMCA,SZ} where it was found that
indeed vortices are generated at velocities above the critical one
about $\sim0.45 c_s$ for $d=10\xi$.

Since the radius of vortex rings (or distance between vortices in
pairs in two dimensions) is about the obstacle size $d$, this
mechanism of vortices emission becomes ineffective for $d\sim\xi$,
and for such small bodies (``impurities'') the critical velocity
arises due to Cherenkov emission of sound waves and coincides,
hence, with the sound velocity $c_s$ \cite{AP04}. Obviously,
emission of waves in a supersonic flow past an obstacle remains
effective also for large obstacles with $d\gg\xi$, but in this
case emitted waves are not linear sound waves and, instead of a
Cherenkov cone, we arrive at the above mentioned dispersive shock
consisting of oblique spatial solitons. Actually, these shock
waves have been observed in the experiment \cite{private} where
the wave pattern consists of a series of distinct oblique traces
which cannot be attributed to a linear Cherenkov radiation
implying appearance of a single cone but can be explained by
generation of supersonic spatial solitons in the flow similar to
those predicted by the KdV dynamics in \cite{karpman},
\cite{GKKE95}. An easy estimate shows that, after long enough time
of expansion, the velocity of the flow past a body is much greater
than the local sound velocity in BEC near this body. Indeed, the
flow velocity $u$ in a free expansion has the order of magnitude
of the sound velocity at the center of BEC before its release and
it is known that the sound speed in BEC is proportional to the
square root of density. Since for the expansion time
$t\gg\omega_\bot^{-1}$ ($\omega_\bot$ is the radial trap frequency
before expansion) the density is proportional to $t^{-2}$ (see,
e.g., \cite{kamch04}), we find that the ratio of the expansion
flow velocity to the local sound speed, i.e. the Mach number $M$,
is about
\begin{equation}\label{eq3}
   M\sim \omega_\bot t\gg 1
\end{equation}
for $t\gg\omega_\bot^{-1}$. This corresponds approximately to
conditions of the experiment \cite{private} where formation of a
``fan'' of spatial solitons was observed. Motivated by the results
of this experiment and above physical argumentation, we shall
develop here the theory of hypersonic spatial dispersive shocks on
the basis of the GP equation (\ref{eq1}).

In accordance with the experiment \cite{exp,private}, we consider
a two-dimensional flow of the condensate, so that the condensate
wave function $\psi$ depends on only two spatial coordinates
$\br=(x,y)$. To simplify the theory, we assume that the
characteristic size of the body is much less than its distance
from the center of the trap, so that incoming flow can be
considered as uniform with constant density $n_0$ of atoms and
constant velocity ${\bf u}_0$ directed parallel to $x$ axis. It is
convenient to transform Eq.~(\ref{eq1}) to a ``hydrodynamic'' form
by means of the substitution
\begin{equation}\label{eq4}
   \psi(\br,t)=\sqrt{n(\br,t)}\exp\left(\frac{i}{\hbar}
   \int^{\br}{\bf u}(\br',t)d\br'\right),
\end{equation}
where $n(\br,t)$ is density of atoms in BEC and ${\bf u}(\br,t)$
denotes its velocity field, and to introduce dimensionless variables
$
   \tilde{x}=x/\sqrt{2}\xi,\, \tilde{y}=y/\sqrt{2}\xi,\,
   \tilde{t}=(c_s/2\sqrt{2}\xi)t,\,
   \tilde{n}=n/n_0,\, \tilde{\bf u}={\bf u}/c_s,
$
where numerical constants are introduced for future convenience. As a result of
this transformation we obtain the system (we omit tildes for convenience
of the notation)
\begin{equation}\label{eq8}
\begin{split}
   \tfrac12 n_t+\nabla(n\bu)=0,\\
   \tfrac12 \bu_t+(\bu\nabla)\bu+\nabla n+\nabla\left[\frac{(\nabla n)^2}{8n^2}
   -\frac{\Delta n}{4n}\right]=0
   \end{split}
\end{equation}
(where $\nabla=(\prt_x,\prt_y)$) which should be solved with the boundary
conditions
\begin{equation}\label{eq10}
   n=1,\quad \bu=(M,0)\quad\text{at}\quad x\to-\infty
\end{equation}
for incoming flow and
\begin{equation}\label{eq11}
   \left.\bu\cdot{\bf N}\right|_S=0
\end{equation}
at the body surface $S$, where ${\bf N}$ denotes a unit vector of
outer normal to the surface $S$. Under our assumption that the
size of the body is much less than the distance from the center of
the cylindrically symmetrical flow, the characteristics of the
flow near the body change with the time very slowly, so that the
arising  wave pattern can be considered as quasi-stationary.
Hence, we can confine ourselves to the steady solutions of our
problem (\ref{eq8})--(\ref{eq11}) and replace Eqs.~(\ref{eq8}) by
their stationary versions for two-dimensional stationary velocity
field $\bu=(u(x,y),v(x,y))$:
\begin{equation}\label{eq12}
\begin{split}
   (nu)_x+(nv)_y=0,\\
   uu_x+vu_y+n_x+\left(\frac{n_x^2+n_y^2}{8n^2}-
   \frac{n_{xx}+n_{yy}}{4n}\right)_x=0,\\
   uv_x+vv_y+n_y+\left(\frac{n_x^2+n_y^2}{8n^2}-
   \frac{n_{xx}+n_{yy}}{4n}\right)_y=0.
   \end{split}
\end{equation}
If the body is symmetric with respect to $x$ axis and the form of
its boundary is given by the function $y=\pm F(x)$, $F(0)=0,$
$F(L)=0$, $L$ being the longitudinal size of the body in
dimensionless units, then we can confine ourselves to consideration
of only the upper half-plane $y>0$, so that ${\bf
N}\propto(F'(x),-1)$, and the boundary conditions (\ref{eq10}),
(\ref{eq11}) are transformed to
\begin{equation}\label{eq14a}
   n=1,\quad u=M,\quad v=0\quad\text{at}\quad x\to-\infty,
\end{equation}
\begin{equation}\label{eq14b}
   v=uF'(x)\quad\text{at}\quad y=F(x).
\end{equation}
The system  (\ref{eq12})--(\ref{eq14b}) is still too complicated
for analytical treatment. However, the flow is supposed to be
supersonic (see (\ref{eq3})) which allows us to asymptotically
transform Eqs.~(\ref{eq12})--(\ref{eq14b}) to a much simpler
problem of 1D ``unsteady'' flow along $y$ axis with the scaled $x$
coordinate playing the role of ``time'' \cite{ekt04}. To this end,
we substitute into Eqs.~(\ref{eq12}) new variables
\begin{equation}\label{eq15}
   u=M+u_1+O(1/M),\quad T=x/2M,\quad Y=y,
\end{equation}
assuming $M^{-1} \ll 1$. Then to leading order we obtain
\begin{equation}\label{eq16}
\begin{split}
   \tfrac12 n_T+(nv)_Y=0,\\
   \tfrac12 v_T+vv_Y+n_Y+\left(\frac{n_Y^2}{8n^2}
   -\frac{n_{YY}}{4n}\right)_Y=0,
   \end{split}
\end{equation}
\begin{equation}\label{eq18}
   \tfrac12 u_{1T}+vu_{1Y}=0.
\end{equation}
Equations (\ref{eq16}) represent the hydrodynamic form of 1D
defocusing nonlinear Schr\"odinger (NLS) equation
\begin{equation}\label{nls-1D}
   i\Psi_T+\Psi_{YY}-2|\Psi|^2\Psi=0
\end{equation}
for a complex field variable
\begin{equation}
\Psi=\sqrt{n}\exp\left(i\int^Y v(Y',t)dY'\right),
\end{equation}
 and we can apply
well-known methods to their study.  If $n$ and $v$ are found from
the system (\ref{eq16}), then the correction $u_1$ to the $x$
component of velocity can be calculated with the use of
Eq.~(\ref{eq18}). It is remarkable that in the case of a slender
body, for which $M \alpha \lesssim 1$ where
$\alpha=\mathrm{max}|F'(x)|$, the boundary condition (\ref{eq14b})
reduces (to leading order in $M^{-1}$) to
\begin{equation}\label{eq19}
   v=v_p=\tfrac12 df/dT\quad\text{at}\quad Y=f(T),
\end{equation}
where $f(T)=F(2 MT)$,
and it does not contain the $u$-variable in this approximation.

Thus, we have reduced the problem of flow past a slender body to
the classical ``piston'' problem for 1D flow along a tube with a
piston moving according to the law (\ref{eq19}). In ordinary gas
dynamics, such a piston motion leads to the generation of two
shock waves which form due to the breaking of the flow profile
during two different phases of the piston motion: forward and
reverse. In the original $2D$ problem this corresponds to the
supersonic flow past the front and the rear parts of the body
respectively and is accompanied by the occurrence of two spatial
shocks (oblique compression jumps) spreading from the body edges
(see \cite{LL6} for instance).

In contrast to the classical gas dynamics, the piston problem is
now posed for dispersive equations (\ref{eq16}).  Before we
proceed with the analysis of this problem we outline the
qualitative structure of the dispersive flow past finite-length
body using the results in \cite{karpman}, \cite{GKKE95}. This will
help us then to reformulate the piston problem for the NLS
equation in terms of much better explored initial-value problem.

In dispersive hydrodynamics, both shocks (breaking singularities)
spreading from the body edges resolve into nonlinear oscillatory
zones,  {\it the dispersive shocks}. At finite distances from the
body surface these two dispersive shocks have similar structure
(see \cite{GP73}): each represents a modulated nonlinear wave
having a form close to a chain of solitons at one edge of the
oscillatory zone and degenerating into a linear wave at the
opposite edge. However, at large distances from the body the two
dispersive shocks demonstrate drastically different behaviour: in
the physical systems with negative dispersion studied in
\cite{GKKE95} the dispersive shock spreading from the front edge
of the body transforms into a soliton train while the dispersive
shock forming at the rear end of the body completely degenerates
into a small-amplitude linear radiation. In the case of the NLS
equation describing nonlinear waves with {\it positive dispersion}
the qualitative picture will be reversed, i.e. the ``nontrivial''
dispersive shock transforming into a (dark) soliton train will
form due to the flow past the rear part of the body (reverse
motion of the piston in the corresponding nonstationary problem).
\begin{figure}[ht]
\centerline{\includegraphics[width=7cm,height=
7cm,clip]{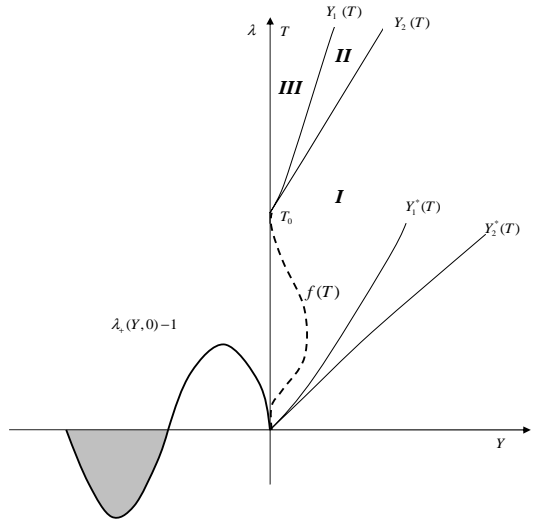}}\vspace{0.3 true cm} \caption{$(Y,T)$-plane of
the piston problem ($Y>0$) and the equivalent initial data
$\lambda_+(Y,0))$ ($Y<0$) for the Riemann invariant. The shaded
area marks a ``soliton'' part of the initial profile. Dashed line:
the ``piston'' trajectory $Y=f(T)$. The lines $Y^*_{1,2}(T)$,
$Y_{1,2}(T)$ are the edges of the front and rear dispersive shocks
respectively.} \label{fig1}
\end{figure}
In the present work we will be concerned with this wave only.

We now formulate the problem of the rear dispersive shock
description in a mathematically consistent way. The relevant part
of the $(Y,T)$-plane in the auxiliary piston problem is subdivided
into three distinct regions (see Fig.~1).
Generally, for $Y>f(T)$, the ``gas'' is put into motion by the
``piston'' moving according to Eq.~(\ref{eq19}) and in the region
$I$ near the ``piston'' the gas motion can be described by the
dispersionless limit
\begin{equation}\label{eq21}
   \tfrac12 n_T+(nv)_Y=0,\quad \tfrac12 v_T+vv_T+n_Y=0
\end{equation}
of Eqs.~(\ref{eq16}). But formal solution of the dispersionless
equations cannot be extended to the whole $(Y,T)$-plane because
the $Y$-derivatives blow up along some line in this plane. Hence,
here we have to take into account the dispersion effects which
lead to formation of the region $II$ filled with  nonlinear
oscillations---a dispersive shock. Close to its boundary
$Y=Y_1(T)$ the oscillations tend to a train of dark solitons of
the NLS equation (\ref{nls-1D}), and at the opposite boundary
$Y=Y_2(T)$ the amplitude of oscillations tends to zero which
corresponds to a linear ``sound'' wave propagating into the
undisturbed region $III$ with $Y>Y_2(T)$. For $T \gg L/M$ the
whole structure can be asymptotically represented as a soliton
train \cite{GKKE95}. Thus, in the case of a macroscopic obstacle
with characteristic size much greater than the healing length
$\xi$ the Cherenkov cone ``unfolds'' into a ``fan'' of solitons.

The outlined nonlinear dispersive flow is described most
conveniently in terms of the Riemann invariants (see, e.g., \cite{
kamch2000, kku02}). In the regions $I$ and $III$ of the smooth
flow these are the Riemann invariants of the dispersionless system
(\ref{eq21})
\begin{equation}\label{eq20}
   \la_\pm=\tfrac12{v}\pm\sqrt{n} \, ,
\end{equation}
which satisfy the equations
\begin{equation}\label{er}
\frac{\partial \la_{\pm}}{\partial T} + V_{\pm}(\la_+, \la_-)
\frac{\partial \la_{\pm}}{\partial Y}=0\, ,
\end{equation}
where the characteristic velocities are
\begin{equation}\label{V}
V_+={3 \la_+}+ {\la_-} \, ,
\qquad
V_-={3\la_-}+ {\la_+} \, .
\end{equation} In the dispersive shock region
$II$,  there are four Riemann invariants $\la_i,$ $i=1,2,3,4,$
which parameterize the modulated periodic solution of the full
system (\ref{eq16}), and obey the corresponding Whitham modulation
equations \cite{wh}
\begin{equation}\label{wh}
\frac{\partial \la_i}{\partial T}+V_i(\la_1, \la_2,\la_3, \la_4)
\frac{\partial \la_i}{\partial Y}=0 \, ,\quad i=1,2,3,4 .
\end{equation}
For the case of the defocusing NLS equation the characteristic
velocities $V_i(\la_1, \dots, \la_4)$ represent certain
combinations of the complete elliptic integrals of the first and
the second kind \cite{pavlov} (see also \cite{ kamch2000, kku02}).
We do not need the concrete expressions for them here.

The dispersionless Riemann invariants (\ref{eq20}) are constant
along the corresponding families of characteristics of the system
(\ref{eq21}) and they must satisfy the matching conditions with
two of the Riemann invariants $\la_i$ of the Whitham equations
(\ref{wh}) along the lines $Y=Y_{1,2}(T)$, which represent (free)
boundaries of the dispersive shock (see \cite{GK87, EK95} for a
detailed description of the matching problem for the solutions of
systems (\ref{er}) and (\ref{wh})).

In the undisturbed region $III$, where $v=0,$ $n=1$, both Riemann
invariants (\ref{eq20}) are constant: $\la_\pm=\pm1$. According to
the well-known argumentation about the flow adjacent to a simple
wave (see \cite{LL6}, Section 104), one of the Riemann invariants
$\la_i,$ $i=1,2,3,4,$ which matches with, say, $\la_-$ (for sake
of definiteness we assume that it is $\la_4$) must also be
constant in the whole region $II$: $\la_4=\la_-=-1$ in $II$. On
the other hand, the gas motion in the region $I$ produced by the
``piston'' is described by a simple wave solution of
Eqs.~(\ref{eq21}) (see \cite{LL6}, Section 101) again with one
Riemann invariant constant everywhere in $I$. It must match with
$\la_4$ along the characteristic line $Y=Y_1(T)$ so that
$\la_-=\la_4=-1$ in the whole $(Y,T)$-plane including the
trajectory of the ``piston''. Hence, we have at the ``piston''
$v_p/2-\sqrt{n_p}=-1$ which yields the gas density,
$
   n_p=(v_p+2)^2/4,
$
and, consequently, the values of both Riemann invariants:
\begin{equation}\label{eq23}
   \la_-^p=-1,\quad \la_+^p=\tfrac12
   df/dT+1\quad\text{at}\quad Y=f(T).
\end{equation}
To use the method of Ref.~\cite{kku02}, we have to transform these
boundary conditions at the ``piston'' to the equivalent initial
conditions at $T=0,\,Y<0$ (see Fig.~1). This  problem for the
system (\ref{eq21}) can  be easily solved using characteristics.
Indeed, we have $\la_-=-1$, hence $\la_+$ obeys the simple wave
equation following from (\ref{er}) (see, e.g., \cite{kku02})
\begin{equation}\label{eq24}
   \frac{\prt\la_+}{\prt T}+{(3\la_+-1)}\frac{\prt\la_+}{\prt Y}=0
\end{equation}
whose general solution is
\begin{equation}
   Y=(3\la_+-1)T+\overline{Y}(\la_+) \, .
\end{equation}
The function $\overline{Y}(\la_+)$ must be chosen so that the
condition (\ref{eq23}) is satisfied. This gives at once
\begin{equation}
\overline{Y}(\la_+)=f(\tau)-(3\la_+-1)\tau,
\end{equation}
where $\tau$ is determined implicitly by the equation
$\la_+=\tfrac12 f'(\tau)+1$. Thus, we arrive at a parametric form
of the equivalent initial distribution of the Riemann invariant
$\la_+$:
\begin{equation}\label{eq27}
   \la_+=\tfrac12 f'(\tau)+1,\quad Y=f(\tau)-(\tfrac32 f'(\tau)+2)\tau.
\end{equation}
In principle, knowing the initial data for the Riemann invariants
$\la_{\pm}$, one can construct the full solution of the Whitham
system (\ref{wh}) using the extended Gurevich-Pitaevskii problem
formulation (see \cite{GK87},\cite{EK95}) and thus, to
asymptotically describe the dispersive shock for all $Y, T$.
However, if one is interested in the asymptotic structure of the
flow in the region far enough from the body where the rear
dispersive shock develops into a ``fan'' of spatial solitons well
separated from each other, one can take advantage of a more simple
asymptotic method of Ref.~\cite{kku02}.

According to the asymptotic method developed in Ref.~\cite{kku02}
each soliton in the soliton train evolving from the initial pulse
is parameterized by the eigenvalue $\la_k$ of the generalized
Bohr-Sommerfeld quantization rule
\begin{equation}\label{eq28}
\begin{split}
   \oint\sqrt{(\la_k-\la_+)(\la_k-\la_-)}\,dY=2\pi(k+\tfrac12),\quad
   k=0,1,\ldots,K,
   \end{split}
\end{equation}
where in our case $\la_+(Y)$ is given by Eq.~(\ref{eq27}), $\la_-=-1$, and
the integration is taken over the cycle around two turning points where
the integrand functions vanishes. Returning to the spatial coordinates
(\ref{eq15}), we find the profile of the $\la_k$-soliton in the train as
(see \cite{kku02})
\begin{equation}\label{eq29}
   n_k(x,y)=1-\frac{1-\la_k^2}{\cosh^2[\sqrt{1-\la_k^2}(y-(\la_k/M)x)]},
\end{equation}
that is the ``fan'' of spatial dark solitons in the shock is made of
soliton ``feathers'' lying asymptotically along the lines
\begin{equation}\label{eq30}
   y=({\la_k}/M) x,\quad k=0,1,\ldots,K,
\end{equation}
in the upper half-plane and symmetric ``fan'' of solitons is
generated in the lower half-plane.

Let us illustrate this theory by an example of a body with a parabolic profile
\begin{equation}\label{eq31}
   y=F(x)=\alpha x(L-x)/(2M)^2,\quad 0\leq x\leq L,
\end{equation}
or
\begin{equation}\label{profile}
    Y=\alpha T(T_0-T),\quad 0\leq T\leq T_0,
\end{equation}
where $T_0=L/2M$,
so that the initial condition (\ref{eq27}) takes the form
\begin{equation}\label{eq32}
   \la_+=\tfrac12\alpha(T_0-2\tau)+1,\quad Y=\tau(2\alpha\tau-\alpha T_0/2-2),
\end{equation}
for $0\leq\tau\leq T_0$, that is in the interval
$-T_0(2-3\alpha T_0/2) \leq Y\leq 0$, and $\la_+=1$ outside this
interval; see Fig.~2.
\begin{figure}[ht]
\centerline{\includegraphics[width=7cm,height=
5cm,clip]{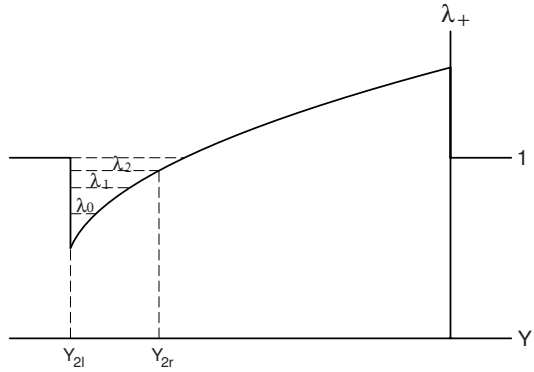}} \vspace{0.3 true cm} \caption{The plot of
$\la_+$ as a function of $Y$ for the parabolic body profile
(\ref{eq31}). Dashed lines labeled by $\la_0,\,\la_1,\,\la_2$ etc
denote the levels corresponding to these eigenvalues, $Y_{2l}$ and
$Y_{2r}$ illustrate positions of the left and right turning points
corresponding to the eigenvalue $\la_2$ in this instance. }
\label{fig2}
\end{figure}
 Its part with $\la_+>1$ evolves into non-solitonic wave which
does not give any contribution into the shock. However, its
``well'' part $\la_+<1$ leads to the bound states in the spectral
problem (\ref{eq28}) and, hence, to the train of spatial solitons
(\ref{eq29}) generated in the shock. Integral in (\ref{eq28}) is
calculated in a closed form which gives the equation
\begin{equation}\label{2-1}
\begin{split}
   \frac{4\sqrt{2}}{15\pi\alpha}(1+4\la_k-\tfrac{27}{4}
   \alpha T_0)(\tfrac12\alpha T_0-1+\la_k)^{3/2}=k+\tfrac12,\quad
    k=0,1,\ldots,K,
   \end{split}
\end{equation}
and its roots $\la_k$ must lie in the interval $1-\alpha T_0/2<
\la_k<1$. The greatest root $\la_K$ has a value close to unity
so that the number of solitons in the shock is given approximately
by (\ref{2-1})
with $\la_k=1$, $k=K$. To transform this estimate to dimensional
variables, we take $l=\sqrt{2}\xi L=2\sqrt{2}\xi MT_0$ as a
longitudinal size  of the obstacle, $d=2\sqrt{2}\xi f(T_0/2)=
\sqrt{2}\xi\alpha T_0^2$ as its transverse size and obtain
\begin{equation}\label{2-2}
   K\cong\frac1{3\pi}\left(1-\frac{27}{10}\frac{Md}l\right)
   \sqrt{\frac{ld}{M\xi^2}}.
\end{equation}
Although this formula is derived for a slender body, we can
use it as a rough estimate of a number of solitons in the shock
generated in the supersonic flow past an obstacle:
\begin{equation}
   K\cong\mathrm{const}\cdot({{ld}/{M\xi^2}})^{1/2},
\end{equation}
where $K$ must be much greater than unity. The most shallow dark
soliton lies near the Cherenkov cone with angle $\theta\cong
1/M=c_s/u_0$ with respect to the direction of the flow. The
resulting pattern is shown in Fig.~3.
\begin{figure}[ht]
\centerline{\includegraphics[width=9cm,height=
5cm,clip]{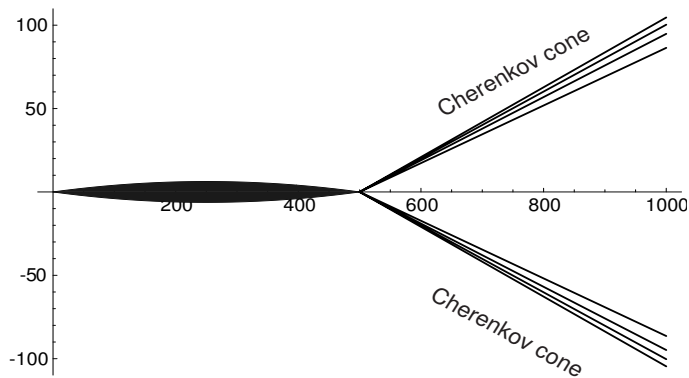}} \vspace{0.3 true cm} \caption{The pattern of
dispersive shocks generated by a BEC supersonic flow with $M=5$
past a slender body (black cigar-shaped figure). The oblique lines
represent the traces of dark solitons in the $(x,y)$-plane. All
dimensions are measured in units of $\sqrt{2}\xi$.} \label{fig3}
\end{figure}

In conclusion, we have developed the theory of spatial dispersive
shock waves generated by a flow of Bose-Einstein condensate past a
slender body. The theory predicts formation of a series of oblique
spatial solitons in the flow and explains qualitatively the shock
patterns observed in the experiment \cite{private}.

We are grateful to E.A. Cornell and P. Engels for information
about the results of the experiments of JILA group prior to
publication. AMK thanks RFBR (grant 05-02-17351) for financial
support.

\end{document}